\documentclass[conference]{IEEEconf}

\usepackage{graphicx}
\usepackage{multirow}
\usepackage{titlesec}
\usepackage{balance} 
\usepackage{stfloats}
\usepackage{xcolor}
\usepackage{enumitem}
\usepackage{xspace}

\usepackage{subcaption}
\usepackage{graphicx}

\usepackage{hyperref}
\hypersetup{hypertexnames=false}

\usepackage[most]{tcolorbox}

\usepackage{wasysym}

\usepackage{booktabs}

\renewcommand{\thesection}{\arabic{section}}
\renewcommand{\thesubsection}{\thesection.\arabic{subsection}}
\renewcommand\thesubsectiondis{\thesection.\arabic{subsection}}% arabic numerals for the subsections

\usepackage[table]{xcolor}
\definecolor{TAM}{RGB}{67, 116, 160}

\newcommand{\naila}{\textsc{naila}\xspace}

\begin{document}

\title{\textbf{\Large Autonomous LLM-generated Feedback for Student Exercises in Introductory Software Engineering Courses\\}}

\author{Andreas Metzger\\
	\normalsize University of Duisburg-Essen, Germany\\
	\normalsize andreas.metzger@paluno.uni-due.de\\
}
%+++++++++++++++++++++++++++++++++++++++++++

%\specialpapernotice{(Invited Paper)}

\maketitle
\begin{abstract}
Introductory Software Engineering (SE) courses face rapidly increasing student enrollment numbers, participants with diverse backgrounds and the influence of Generative AI (GenAI) solutions. 
High teacher-to-student ratios often challenge providing timely, high-quality, and personalized feedback a significant challenge for educators. 
To address these challenges, we introduce NAILA, a tool that provides 24/7 autonomous feedback for student exercises.
Utilizing GenAI in the form of modern LLMs, NAILA processes student solutions provided in open document formats, evaluating them against teacher-defined model solutions through specialized prompt templates.
We conducted an empirical study involving 900+ active students at the University of Duisburg-Essen to assess four main research questions investigating (1) the underlying motivations that drive students to either adopt or reject  NAILA, (2) user acceptance by measuring perceived usefulness and ease of use alongside subjective learning progress, (3) how often and how consistently students engage with NAILA, and (4) how using NAILA to receive AI feedback impacts on academic performance compared to human feedback.
% We conclude by discussing potential future directions for exercise feedback and regulatory considerations for the use of AI in education.

% \todo{turn of paragraph identation before subm}
\end{abstract}
\IEEEoverridecommandlockouts
\vspace{1.5ex}

% for peerreview papers, inserts a page break and creates the second title.
% Will be ignored for other modes.
\IEEEpeerreviewmaketitle

\section{Problem Statement}
\label{sec:intro}

Teaching Software Engineering (SE) programmes -- particularly undergraduate introductory courses on SE -- faces important challenges due to recent trends and developments.
We first discuss and illustrate these challenges using actual data from our first-year undergraduate course on "introduction to SE" at the University of Duisburg-Essen (Germany).
We then then present the key contributions of this paper.

\subsection{Challenges}
\label{sec:chall}
    \textbf{(a) High number of students:} As expressed in~\cite{FanNDR25,LiaoJCS24}, the growing number of students enrolling in computer science programs is pushing educators to their limits. 
    This poses significant challenges to computer science education, and consequently also to 
    SE education.

\begin{figure}[htb]
        \centering
        \includegraphics[width=.9\linewidth]{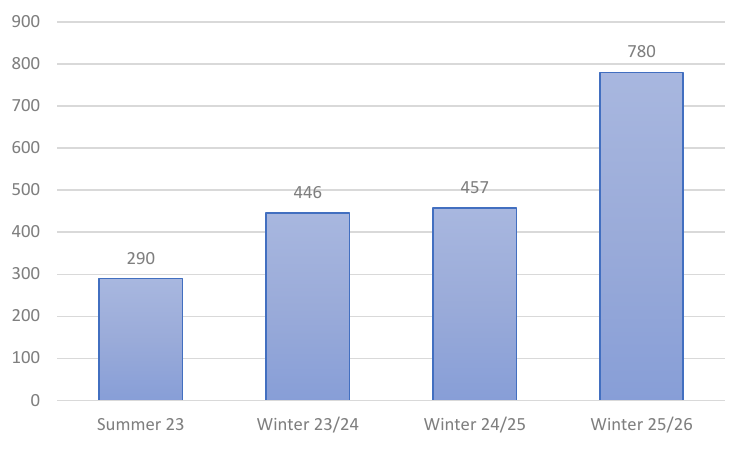}
        \caption{Total number of participants in UDE's introductory SE course (CAGR = 32.7\%)}
        \label{fig:participants}
    \end{figure}

\begin{figure}[htb]
        \centering
        \includegraphics[width=.9\linewidth]{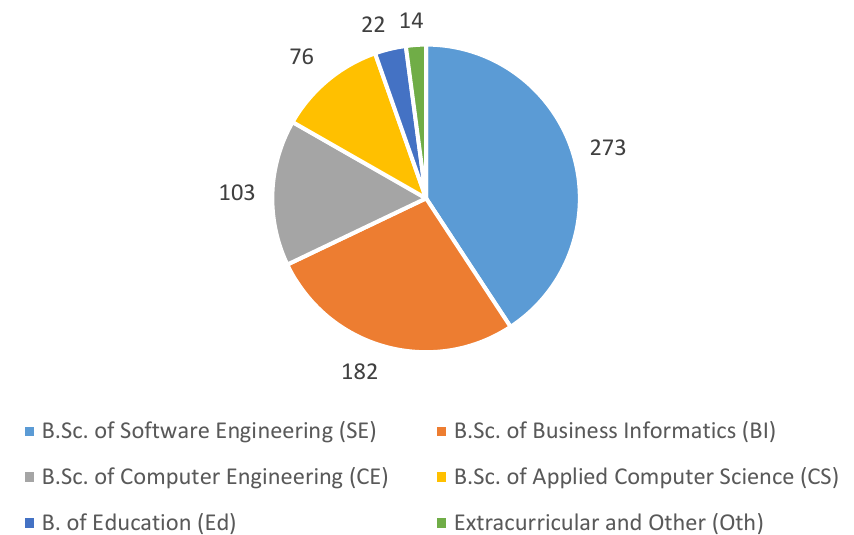}
        \caption{Degree programs in UDE's introductory SE course ($N = 670$)}
        \label{fig:degree_distribution}
    \end{figure}

    Fig~\ref{fig:participants} depicts the students' participation in our undergraduate SE course measured by the number of students that have been admitted to the final exams.
    As can be seen, enrollment numbers drastically increased (with a CAGR of 32.7\%) due to the fact that our SE course became mandatory in five different degree programs (cf. Fig~\ref{fig:degree_distribution}).
   
    Specifically, the resulting high teacher-to-student ratio makes it challenging to provide timely and high-quality feedback~\cite{SDK25}.
    Manual assessment can be prone to subjectivity and inconsistencies across a large number of submissions, and often teachers do not have sufficient time and resources to provide detailed feedback.

    \textbf{(b) Diverse types of students:}
    Introductory courses often include interdisciplinary students who are not SE majors. 

    In our introductory SE course, we teach students from a broad range of degree programs.
    Fig.~\ref{fig:degree_distribution} shows the distribution of students that participated in the final exam of the winter semester 25/26 course.

    % of B.Sc. of SE (48\%), B.Sc. of Business Informatics (32\%), B.Sc. of Applied Computer Science (8\%), B.Sc. of Computer Engineering (6\%), and  B. of Education (3\%) and a mix of others (3\%).
    
    \textbf{(c) Widespread use of Generative AI (GenAI):}
    Many of the classical student assignments and exercises to reinforce and train the material presented during lectures can easily be solved by modern GenAI solutions, such as Large Langage Models (LLMs).
    As students even in their first year of colleague show an increasing proficiency in using GenAI~\cite{Mead24}, these two factors mean that educators face rapidly changing classroom environments and disrupted teaching principles, as solutions for assignments and exercises may be fully generated via GenAI~\cite{VierhauserGAS24}. 
    
\begin{figure}[htb]
   \centering
        \includegraphics[width=1\linewidth]{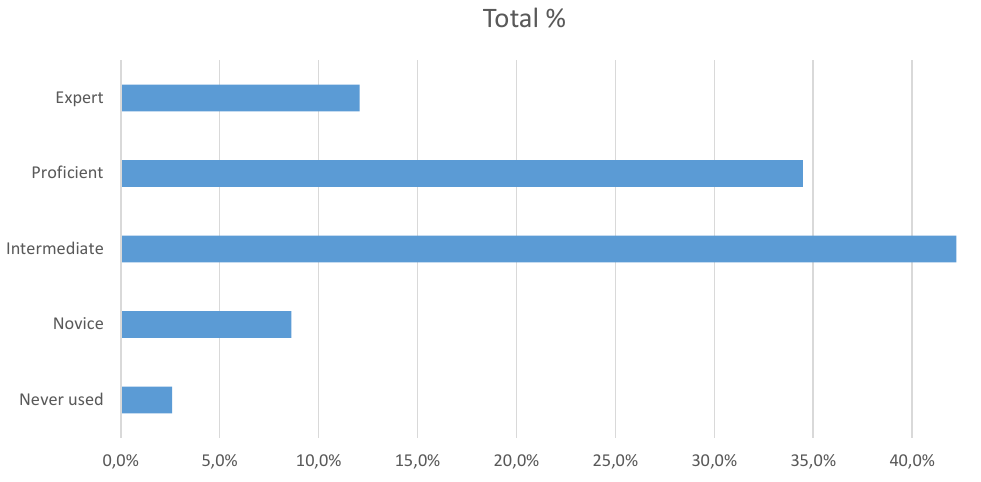}
    \caption{Students AI experience levels in UDE's introductory SE course ($N = 116$; single choice)}
    \label{fig:ai_prof}
    \end{figure}

    In our SE course, the students' self-reported GenAI experience (see Figure~\ref{fig:ai_prof}) indicates that the majority of students (89\%) that responded to an online survey during one of our lectures  has GenAI  knowledge of intermediate or better.

    This means that assignments and exercises -- in their current form -- loose their effectiveness, as it becomes difficult to determine whether the student has learned the given topic~\cite{Datta24}.

    \textbf{(d) Preference for E-Learning:} Survey- and focus-group-based research shows a clear  preference for e-learning, as this provides flexibility for scheduling learning sessions~\cite{DaunBTO25}.
    
    We observe this also in our SE course as visible from Table~\ref{tab:learning_resources}, where videos, AI-based learning and lecture slides are forming the top 3 learning resources.

    \begin{table}[h]
    \centering
    \begin{tabular}{lcc}
    \hline
    \textbf{Learning Resource} & \textbf{Percentage} \\
    \hline
    Explanatory videos  & 81\% \\
    AI-based interactive learning  & 72\% \\
    Lecture slides  & 71\% \\
    E-books / PDF / Kindle  & 37\% \\
    Classic books in paper form  & 25\% \\
    Other  & 11\% \\
    \hline
    \vspace{.01em}
    \end{tabular}
    \caption{The different uses of learning resources as reported by SE students ($N = 359$; multiple choice)}
    \label{tab:learning_resources}
    \end{table}

\vspace{1em}

\subsection{Contributions}
\label{sec:contr}

To address the aforementioned trends and problems, we introduce \naila, the (\underline{N}ovel \underline{AI}-based \underline{L}earning \underline{A}pp).

\naila provides 24/7 autonomous feedback for introductory SE  exercises.
\naila covers a range of different types of SE questions, ranging from ones, where the exact solution is required (i.e., has to match the model solution) to ones where students may answer in a free-text form not prescribed by a model solution.
Thereby \naila goes beyond the typical form of assessment for  programming assignments (see Section~\ref{sec:sota}).
By autonomously providing personalized feedback to students, \naila aids educators in efficiently and effectively delivering high-quality feedback. 

To assess the efficiency and effectiveness of \naila, 
we 
answer  the following research questions:
\begin{itemize}%[label={}]
    \item \textit{RQ1: Why do students use or don’t use NAILA?} This question helps us to understand the motivation for using \naila, but also provides the reasons for why students may have decided against using \naila.
    \item \textit{RQ2: How do students accept and perceive \naila{}?} This question assesses self-reported learning gains of students by determining the perceived usefulness, ease of use, and learning progress when using \naila.
    % typical TAM exercise
    \item \textit{RQ3: How is \naila{} used by students?} This question quantitatively assesses the usage patterns of students, including how many students use \naila and how often and repeatedly for the same exercise they use \naila.
    % both, how many students and how often
    \item \textit{RQ4: How does \naila{} affect learning?} This is the most fundamental research question we aim to answer.
    It assesses in how far \naila's feedback improves the students' academic performance compared to human feedback provided during face-2-face exercise meetings.    
\end{itemize}

The remainder of paper is structured as follows.
Section~\ref{sec:NAILA} describes the architecture and realization of \naila, particularly focusing on how the underlying LLM was prompted.
Section~\ref{sec:empirical} introduces the setup and the results of our empirical study to answer RQ1--RQ4.
Section~\ref{sec:sota} discusses related work.
Section~\ref{sec:discussion} provides a critical discussion how the above challenges and trends are addressed, sketches future directions and raises regulatory considerations
for the use of AI in education.

\section{Realization of \naila}
\label{sec:NAILA}
\begin{figure*}[t]
    \centering
    \includegraphics[width=1\linewidth]{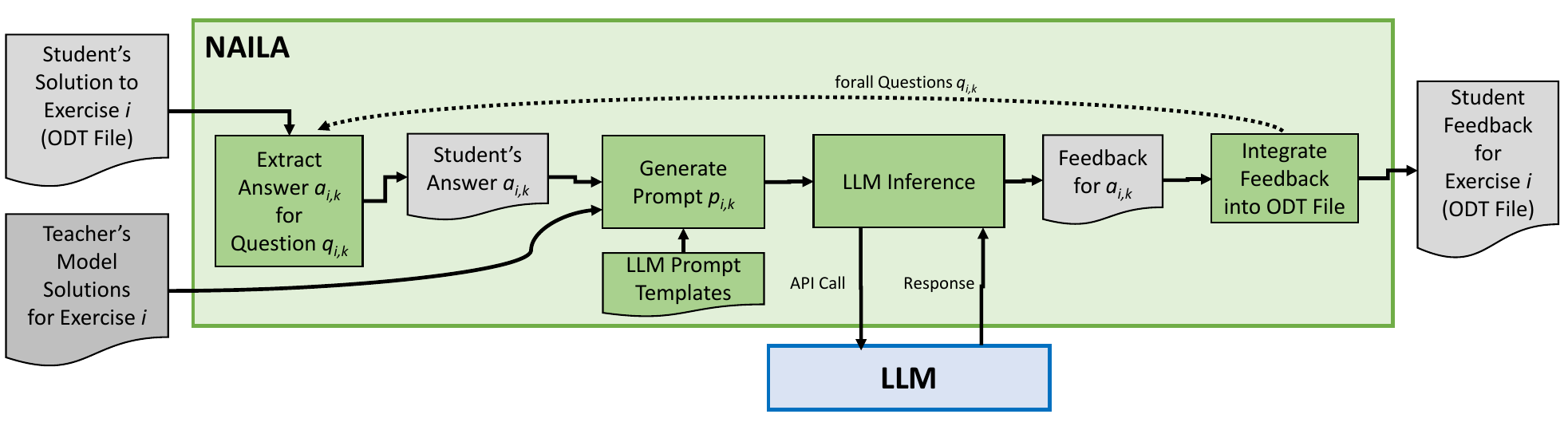}
    \caption{Conceptual Architecture and Data Flow of \naila}
    \label{fig:archi}
\end{figure*}

Figure~\ref{fig:archi} depicts the main components of \naila.
\naila takes a student's solution to a given exercise $i$ and breaks it down into its individual answers $a_{i,k}$ (one per each question $q_{i,k}$).
Exercises are provided as open document (ODT format) files, which  use hashtags '\#' to specify the areas where answers are to be inserted and what the maximum points that can be achieved are.
An example is shown in Figure~\ref{fig:question} below.

\begin{figure}[hbt]
    \centering
\begin{tcolorbox}[colback=gray!10, colframe=gray!90]
\textbf{Question 8.1a)} The following two quality assurance techniques  are available in a software project: \textit{code review}, for which experience shows it finds 10 defects in 5 person hours; \textit{automated unit test and debugging}, which is predicted to find 15 defects, but will require 10 person hours of effort. \textbf{Explain which of the two techniques you would use in a project that has an extremely short deadline.}

\#\# Answer 8.1a Start \#\#  POINTS: 4 \#\#

\textit{\hspace{2em}$<$your answer here$>$}

\#\# Answer 8.1a End \#\#
\end{tcolorbox}

    \caption{Example of an Exercise Question}
    \label{fig:question}
\end{figure}

After extracting the student answer together with the achievable POINTS, \naila creates the actual prompt by filling in the respective type of prompt template and adding the model solution 
followed by the student's answer $a_{i,k}$.
In addition, the prompt instructs the LLM to grade the student solution.

In our current implementation of \naila, we provide three teacher-selectable prompt templates to reflect how stringent the student answer must match the model solution:

\begin{itemize}
    \item \textbf{Close match:} The student solution must cover all the answer elements of the model solution.
    \item \textbf{Partial match:} The student solution must at least cover $n$ of the the answer elements of the model solution.
    \item \textbf{Flexible match:} The model solution is only an example, and the student may answer differently provided that the gist of the answer is maintained and the answer is well argued.
\end{itemize}

The prompt for the above example is shown in Figure~\ref{fig:prompt}.
This prompt is  sent to the LLM via a respective API Call.
In our case, we used Google's Gemini 2.5 Flash, as this model was sufficiently powerful, yet provided answers with a relatively low response time.

\begin{figure}[hbt]
    \centering
\begin{tcolorbox}[colback=gray!10, colframe=gray!90]
\textbf{Prompt for Question 8.1a)}
\vspace{1em}

\footnotesize{\texttt{
<Role> You are a lecturer for the Bachelor's course 'Introduction to Software Engineering' and you are grading and providing feedback for a student's exercise solutions. You may be generous in your assessment, and minor errors or inaccuracies have no impact on the assessment and grading. Spelling and grammar should NOT be corrected!
}}

\footnotesize{\texttt{
<Context> The answer options provided below are all expected in the student solution. The student solution is the solution provided by the student and may contain errors or be incomplete. 
}}

\footnotesize{\texttt{
<Action> Check the extent to which the student solution correctly answers the question regarding the given answer options. The given answer options must be included in a correct and complete student solution. The order of the answers does not matter. Missing spaces do not matter.
}}

\footnotesize{\texttt{
<Output> Award points on a scale from 0 = everything wrong, to 4 = everything correct. Award points with a precision of 0.5 points. Even for an empty submission, award 0 points. Output these points at the very end after the string 'POINTS:' 
}}

\footnotesize{\texttt{
<Question> [see above]
}}

\footnotesize{\texttt{
<Model Answer> While “testing and debugging” is more effective (15 vs 10 defects), one should use "code review”, because it has a higher efficiency (2.0 vs 1.5), meaning it has a higher bug rate per invested hour.
}}

\footnotesize{\texttt{
<Student Solution> [inserted here]
}}
\end{tcolorbox}
    \caption{Prompt for Example Exercise Question}
    \label{fig:prompt}
\end{figure}

Finally, \naila integrates the response received from the LLM for each of the student answers $a_{i,k}$ into the ODT document provided by the student and returns  this ODT document as a download back to the student.

We deployed \naila as a Docker container on the Google Cloud Compute Engine, facilitating elastic computing and storage, while keeping costs to a minimum.
\naila offers a simple user interface, where students can upload their exercise solution after having provided their e-mail address to check whether they are registered for the course.
% Note that the UI is in German, as this is the official language of the course.
% After the file has been processed, it is automatically downloaded.

% \newpage
% \clearpage
\section{Empirical Study}
\label{sec:empirical}

We first provide descriptive statistics on the student population and then report the setup and results for each of our four research questions.

\subsection{Student Population} 
\label{sec:pop}

Of the \textbf{1,140 students} who registered on the learning platform of our SE course (Moodle), \textbf{843 students (74\%) actively participated} in the course.
We counted active participation if a student participated in the introductory test at the beginning of the semester.

The use of \naila was not mandatory, so the students could decide for themselves whether they used AI-based feedback for their exercises or not.
Out of the 843 active students, \textbf{314 (37\%)} opted to use \naila, i.e., we counted them if they used \naila at least once.

\subsection{RQ1: Why do students use or don't use \naila?}
\label{sec:rq1}
\textbf{RQ1 Setup:}
We started by investigating why  students use (\textit{U}) or don't use (\textit{NU}) \naila.
To this end, we performed a voluntary survey among the students, during which we asked the following two groups of multiple-choice questions.

\begin{tcolorbox}[colback=green!20, colframe=black, title= U: Reasons for using \naila (Multiple-Choice)]
% \begin{itemize}
% [label=\footnotesize{\textbf{\arabic*}.}]
\textit{Why did you use \naila?}
\begin{itemize}
    \item \textbf{Deep Understanding: }To better understand complex SE concepts.
    \item \textbf{Practical Application:} To learn how to apply theory to real-world problems.
    \item \textbf{Exam Preparation:} To prepare for upcoming tests and the final exam.
    \item \textbf{Efficiency:} To complete my exercises more quickly and manage my time.
    \item \textbf{Career Skills:} To build technical competencies for my future profession.
    \item \textbf{Other} reasons.
    \end{itemize}
\end{tcolorbox}

\begin{tcolorbox}[colback=red!20, colframe=black, title= NU: Reasons for \underline{not} using \naila (Multiple-Choice)]
\textit{Why did you decide against using \naila?}

\begin{itemize}
% [label=\footnotesize{\textbf{\arabic*}.}]
   \item \textbf{Self-Sufficiency:} The exercises were easy enough to do without help.
    \item \textbf{Non-Participation:} I did not attempt the exercises at all. 
    \item \textbf{Technical Issues:} I experienced bugs or difficulty accessing the tool. 
    \item \textbf{Human Preference:} I preferred getting feedback from instructors or peers. 
    \item \textbf{Low Perceived Value:} I didn't think it would improve my grades or understanding. 
    \item \textbf{Time Constraints:} Using the tool felt like an extra, time-consuming step. 
    \item \textbf{Privacy Concerns:} I was worried about how my data would be used. 
    \item \textbf{Other} reasons.
\end{itemize}
\end{tcolorbox}

\textbf{RQ1 Results:}
A total of 165 students (20\% of active students) participated in the voluntary survey.
Table~\ref{tab:U} shows reported reasons for using \naila.
The main reason (given by ca. 61\% of the students) being to support them in exam preparation.
Interestingly, many students also perceive \naila as a tool to help them gain a better (ca. 32\%)  or more efficient (ca. 24\%) understanding of SE topics.
The practical application of SE concepts was a lesser reason for using \naila (only ca. 21\%).
This can be explained by the fact that the majority of the SE topics presented in the course were of a more fundamental and conceptual nature, and thus many of the exercises were dedicated to discussing and reflecting on these concepts instead of, say, using them to execute some concrete SE task (exceptions being risk prioritization, requirements classification, test case generation, and version control).

\begin{table}[htb]
    \centering
    \begin{tabular}{lcc}
    \hline
\rowcolor{green!20}    \textbf{U} & \textbf{Percentage} \\
    \hline
Exam Preparation	&	60.9\%	\\
Deep Understanding	&	31.5\%	\\
Efficiency	&	23.9\%	\\
Practical Application	&	20.7\%	\\
Career Skills	&	6.5\%	\\
Other	&	5.4\%	\\
    \hline
    \vspace{.01em}
    \end{tabular}
    \caption{Reasons for using NAILA\\($N = 92$; multiple choice)}
    \label{tab:U}
    \end{table}

Table~\ref{tab:NU} shows the reported reasons for \underline{not} using \naila.
One of the top reasons given (ca. 18\%) was that students have not participated in exercises (i.e., neither worked on exercise sheets nor participated  in exercise meetings).
The other most prominent reason (also ca. 18\%) was that they prefer receiving feedback from human instructors rather than from an AI.
Interestingly, all of these students reported an AI proficiency of at least "Intermediate", so having never used AI and thus being skeptical about it, could not have been a reason.

Despite the fact that we offered technical support via online forums, several students (ca. 15\%) reported technical issues.
They suggested that for future courses one could provide a small introductory tutorial to \naila to explain its use and reconsider the use of ODT-Files, which were difficult to edit for some of the students.

\begin{table}[htb]
    \centering
    \begin{tabular}{lcc}
    \hline
\rowcolor{red!20}    \textbf{NU} & \textbf{Percentage} \\
    \hline
Non-Participation in Exercises	&	17.8\%	\\
Human Preference	&	17.8\%	\\
Time Constraints	&	15.1\%	\\
Technical Issues	&	15.1\%	\\
Self-Sufficiency	&	12.3\%	\\
Low Perceived Value	&	12.3\%	\\
Other	&	11.0\%	\\
Privacy Concerns	&	4.1\%	\\
    \hline
    \vspace{.01em}
    \end{tabular}
    \caption{Reasons for not using NAILA\\($N = 73$; multiple choice)}
    \label{tab:NU}
    \end{table}

\subsection{RQ2: How do students assess their learning process using \naila{}?} 
\label{sec:rq2}
\textbf{RQ2 Setup:}
Self-report measures are essential sources of information about how students perceive their learning process~\cite{hadwin2025advancing}.
We therefore performed a voluntary survey in which students are asked dedicated questions to assess their learning process and their learning gains using \naila.

The survey questions are based on the Technology Acceptance Model (TAM)~\cite{davis1989perceived}, which we chose due to its low complexity, empirically validated measurement scales, as well as its robustness~\cite{king2006meta, morris1997user}.
Research demonstrated that TAM can be used to evaluate software prototypes~\cite{morris1997user, laitenberger1998evaluating, davis2004toward}, such as \naila.
While TAM is robust for predicting technology adoption and user satisfaction, it relies on self-reported data.

To effectively collect reports about the students' learning gains, we used the two core dimensions of TAM -- perceived ease of use (PEOU) and perceived usefulness (PU) -- and extended them with a third dimension called perceived learning (PL) as proposed in the literature~\cite{liao2022integration}.
% This is due to the fact that while standard TAM questions are suitable for capturing the students' "belief" of learning, they may not reflect actual cognitive development or knowledge retention.
We understand that  \textit{PL} measures subjective confidence and perceived self-efficacy rather than objective knowledge retention. 
Therefore, RQ2 serves to evaluate the student experience, which we later contrast with actual pedagogical outcomes in RQ4.

Answers are given on a five-point Likert scale: (1) strongly \underline{dis}agree, (2) \underline{dis}agree, (3) undecided, (4) agree, (5) strongly agree.

The following question \textit{PEOU} provides insights about \naila's usability, i.e., how easy it was to use \naila to receive feedback during learning.

\begin{tcolorbox}[colback=TAM!20, colframe=black, title= \textit{PEOU}: Perceived Ease of Use (Likert Scale)]
\textit{How do you perceive the ease of use of \naila?}
\begin{enumerate}
\item It was easy to get feedback from \naila.
\item I learned to use \naila very quickly.
\item The feedback was easy to understand.
\item The feedback was always factually correct.
\item Overall, \naila is user-friendly.

\end{enumerate}
\end{tcolorbox}

The following question \textit{PU}  provides insights into how helpful students perceive the feedback provided by \naila and thereby indicates how students' assessed how \naila did help them learn.

\begin{tcolorbox}[colback=TAM!20, colframe=black, title=\textit{PU}: Perceived Usefulness (Likert Scale)]
\textit{How do you perceive the usefulness of \naila?}
\begin{enumerate}

\item \naila helped me understand course materials more deeply.
\item I finished my learning tasks faster using \naila.
\item \naila helped me produce higher-quality (more accurate) work.
\item Using \naila improved my performance on mid-semester tests.
\item \naila’s feedback was as useful as the feedback from human tutors.
% \item \naila’s feedback was as useful as feedback from human teachers.
\item Overall, \naila helped me to learn.

\end{enumerate}
\end{tcolorbox}

The following question \textit{PL}  aims to understand the learning gains of the students.

\begin{tcolorbox}[colback=TAM!20, colframe=black, title=\textit{PL}: Perceived Learning (Likert Scale)]
\textit{How do you perceive your learning progress using \naila?}
\begin{enumerate}
    \item I feel more confident explaining SE topics because of \naila.
    \item \naila helped me develop practical SE skills. 
    \item I believe I will remember what I learned through \naila for a longer time. 
    \item \naila encouraged me to think more critically about SE topics. 
    \item Overall, \naila helped me to learn better. 
\end{enumerate}
\end{tcolorbox}

To provide a summary of results, we follow the common approach to calculate the average results $\hat{X}$ for  all items $X_i$ within a dimension $X \in $ \{\textit{PEOU}, \textit{PU}, \textit{PL}\}:

$$\hat{X} = \frac{\sum_{i=1}^{n} X_i}{n}$$

We then map $\hat{X}$ to the closest integer number representing a point in the Likert scale.

\begin{figure}[htb]
   \centering
    \includegraphics[width=.7\linewidth]{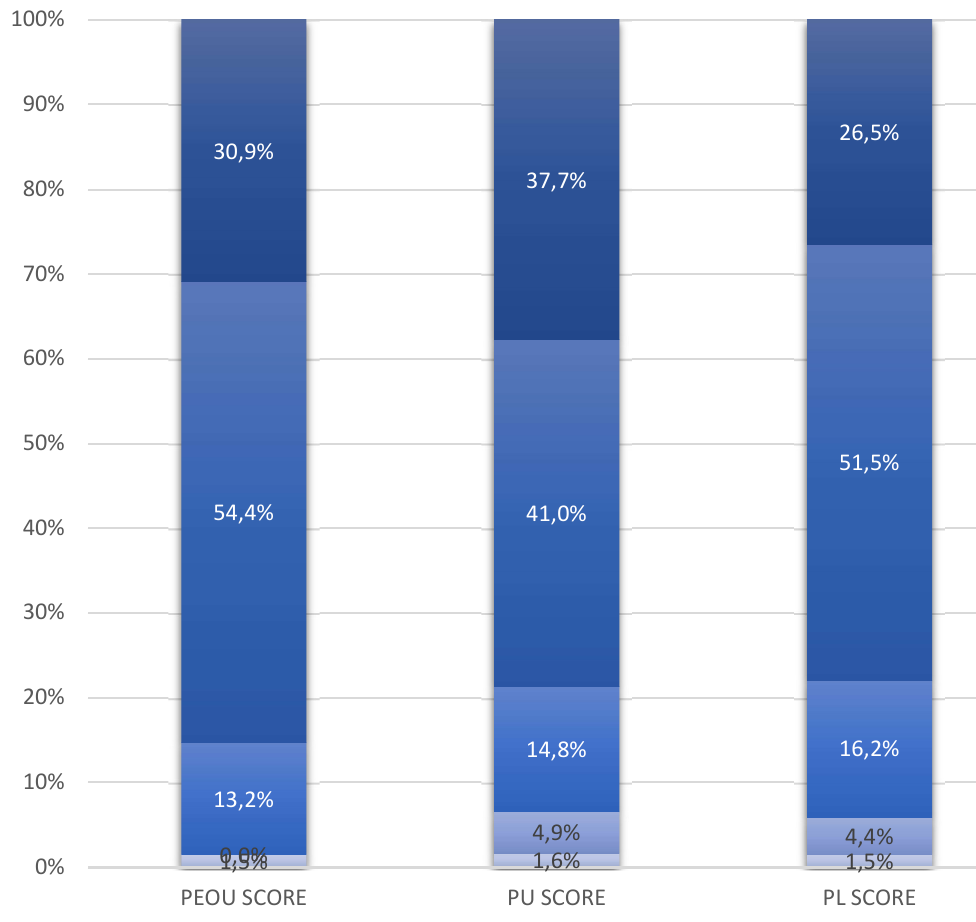}
    \includegraphics[width=1\linewidth]{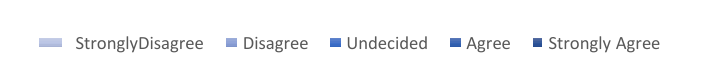}
    \caption{Summary of RQ2 Results\\ ($N = 68$)}
    \label{fig:summaryRQ2}
\end{figure}

\begin{figure}[htb]
  \centering
     \begin{subfigure}[b]{1\linewidth}
         \centering
         \includegraphics[width=\textwidth]{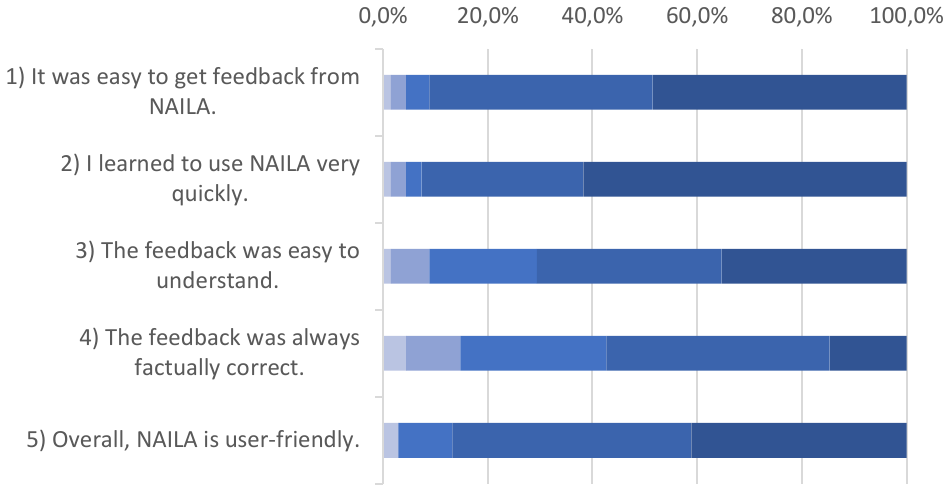}
         \caption{Perceived Ease of Use (\textit{PEOU})}
     \end{subfigure}

     \begin{subfigure}[b]{1\linewidth}
         \centering
         \includegraphics[width=\textwidth]{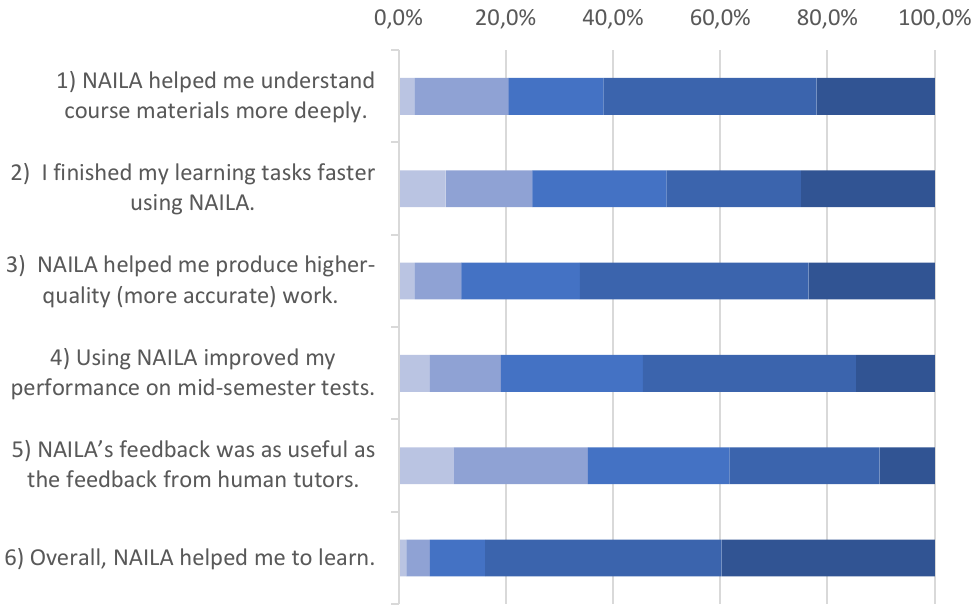}
         \caption{Perceived Usefulness (\textit{PU})}
     \end{subfigure}

     \begin{subfigure}[b]{1\linewidth}
         \centering
         \includegraphics[width=\textwidth]{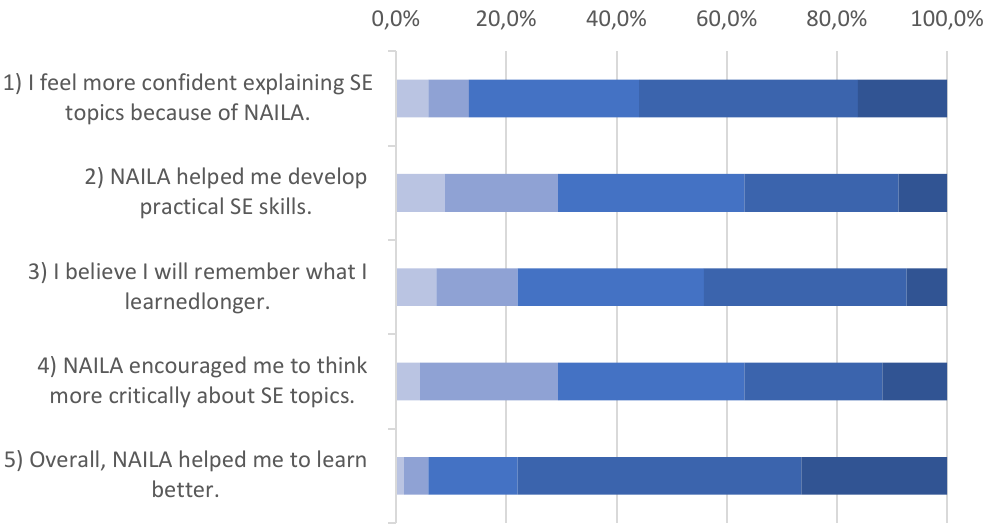}
         \caption{Perceived Learning (\textit{PL})}
     \end{subfigure}

    \includegraphics[width=1\linewidth]{fig/Likert.pdf}
    \caption{Breakdown of RQ2 Results\\ ($N = 68$)}
    \label{fig:breakdown}
\end{figure}

\textbf{RQ2 Results:}
Figure~\ref{fig:summaryRQ2} shows the summarized results of the 68 respondents (74\% of \naila users) for each of the three dimensions, while Figure~\ref{fig:breakdown} provides a breakdown for the individual questions.

The darker the colors are, the better and higher were the benefits and impact of \naila.
As can be seen, in all three dimensions, \naila achieved very promising results with weighted averages of $PEOU = 4.1$, $PU = 4.1$, and $PL = 4.0$, which represents an "agree" on the Likert scale.

Given the high \textit{PEOU}, \textit{PU}, and \textit{PL} scores, we can deduce that students perceived \naila  as a highly beneficial support tool that significantly increased their subjective confidence in the material. 
To determine whether this high self-efficacy translated into actual knowledge retention, these survey results must be interpreted alongside the objective exam performance data evaluated in RQ4.

Given a high \textit{PEOU}, \textit{PU} and \textit{LG}, we can deduce that \naila was perceived by students as a successful tool for learning.
One interesting observation can be made for question \textit{PU}-5.
For this question, \naila received the lowest rating (when compared to the other questions), indicating that 35\% of the students perceived human feedback to be more useful (possibly relating to the "human preference" as observed in RQ1 for \naila non-users).

% \todo{Reliability Check: Run a Cronbach’s Alpha test. For TAM dimensions, you generally want a value $\alpha > 0.70$. This confirms that the items in your PU category are actually measuring the same thing.}

\subsection{RQ3: How is \naila{} used by students?}
\label{sec:rq3}
\textbf{RQ3 Setup:}
Our aim is to assess \naila usage as an indicator for the students' intrinsic motivation and learning engagement~\cite{an2024relationship}.

To determine how students have used \naila, we collected usage logs that recorded the order  when accessing \naila, the number of the exercise that was uploaded for feedback, a unique identifier of the student, as well as the points achieved in the respective exercise iteration.

We quantify \naila usage of (\textit{NU}) of a student by introducing two distinct metrics.
\textit{NUC (Confirm)} captures how well  students performed when deciding they don't need more than one round of AI feedback for an exercise.
\textit{NUR (Remedy)} captures how much students' exercise performance improves when they repetitively use AI feedback for an exercise.
These metrics are computed as follows.

\begin{tcolorbox}[colback=yellow!20, colframe=black, title= \textit{NU}: \naila Usage (numeric scale)]

Let $E_{s,1}$ be the total number of exercises the student $s$ only submitted once, and $E_{s,1+}$ the total number of exercises submitted multiple times.
Let $I_i$ be the initial score and $F_i$ the final score for exercise $i$, then:

$$NUC_s = \frac{1}{|E_{s,1}|} \sum_{i \in E_{s,1}} I_i$$

$$NUR_s = \frac{1}{|E_{s,1+}|} \sum_{i \in E_{s,1+}} RLG_i$$

$$RLG_i = \frac{F_i - I_i}{100\% - I_i}$$

\end{tcolorbox}

\textit{RLG} captures the relative learning gain.
It measures how much a student improved relative to the room a student had to improve.
For example, if a student scores 80\% on the first try and iterates to 100\%, the student improved by 20 points, but captured 100\% of  potential growth ($RLG = 1.0$). 
If a student starts at 40\% and iterates to 60\%, the student also improved by 20 points, but only captured 33\% of potential growth ($RLG = 0.33$). 
This levels the playing field between high-achievers and struggling students.

% Note: For the "ace" student who scores 100% on the first try, the denominator is 0, safely excluding them from this specific metric since they didn't need the tool.

\textbf{RQ3 Results}
Overall 314 students (= 37\% of  actively participating students) used \naila to get feedback for their written exercises.
Of those, 139 (44\%) only used "confirmatory" feedback (\textit{NUC}), 20 (6\%) always used "remedial" feedback (\textit{NUR}), and the remaining 155 (49\%) used both.

\begin{figure}[htb]
  \centering
     \begin{subfigure}[b]{.45\linewidth}
         \centering
         \includegraphics[width=\textwidth]{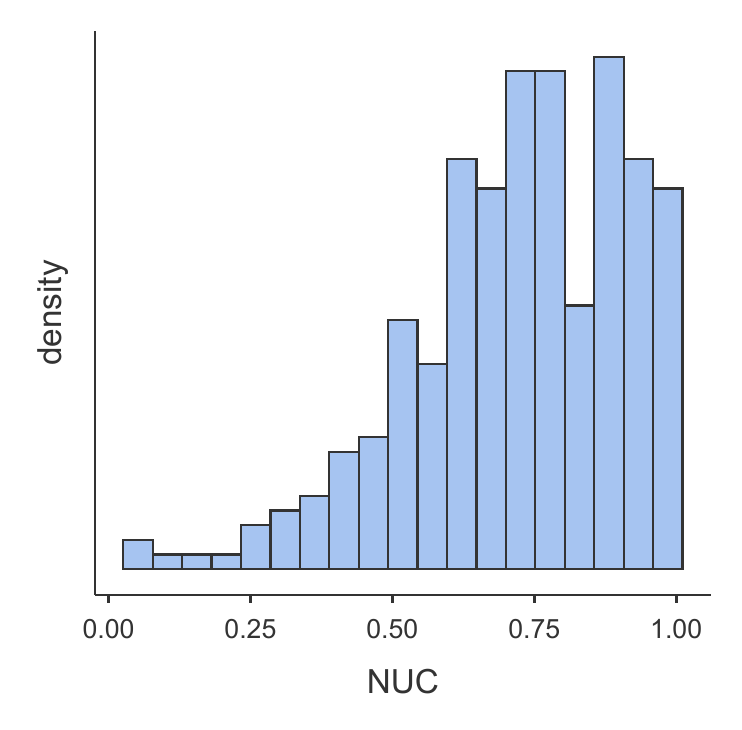}
         % \caption{Perceived Ease of Use (\todo{$N = 57$})}
         % \label{fig:peou-bars}
     \end{subfigure}
     \begin{subfigure}[b]{.45\linewidth}
         \centering
         \includegraphics[width=\textwidth]{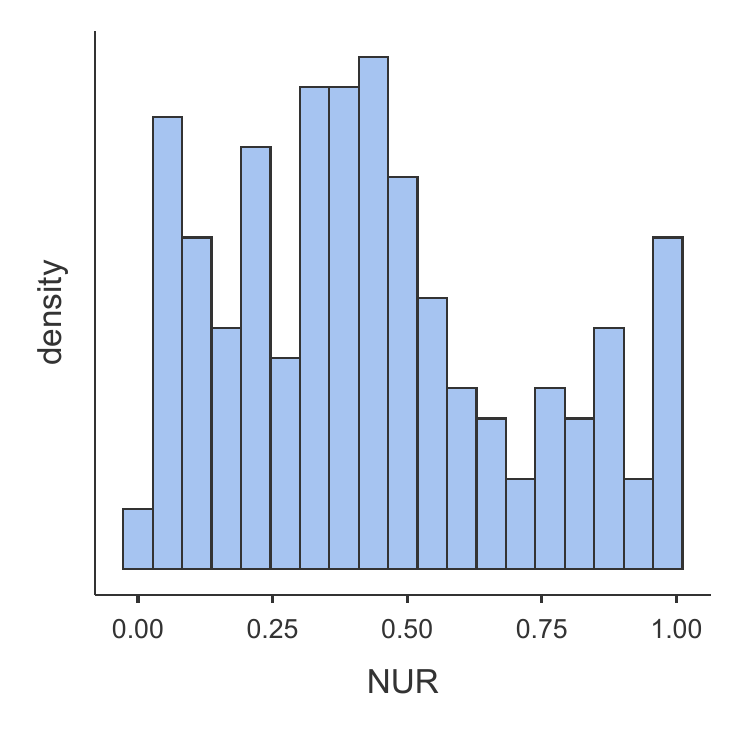}
         % \caption{Perceived Ease of Use (\todo{$N = 57$})}
         % \label{fig:peou-bars}
     \end{subfigure}

    \caption{Histograms of Students' \naila usage\\ ($N_{NUC} = 294, N_{NUR} = 175 $)}
    \label{fig:NU}
\end{figure}

Fig~\ref{fig:NU} shows the histograms for the different \naila usages indicating that in case of \textit{NUC} the majority of students achieved relatively high scores for their first submission, thus providing them with positive reinforcement about their learning process.

% \todo{CHECK: higher NUC or NUR should correlate with higher Perceived Learning (PL) scores from RQ2.}

\subsection{RQ4: How does \naila{} affect learning?} 
\label{sec:rq4}
\textbf{RQ4 Setup:}
To achieve a quantitative assessment of how \naila may affect learning, we measured the students' performance in the exam taken at the end of the semester (\textit{SP}).
The exam took 60 minutes and consisted of multiple-choice questions (ca. 30\%) and free-text questions. 

% Given the short time frame between the date of the exam and the paper submission deadline, we perform a fully automated correction of the exam (e.g., as also suggested by~\cite{Bassner2025}).
% This entails a syntactic check of the multiple-choice questions, and LLM-based correction of the free-text answers by adapting the tooling presented in Section~\ref{sec:NAILA} to the exam setting.
% To test the quality of the LLM-based correction, we checked a sample of exams and answers.

As the use of \naila was optional and therefore students could decide for themselves whether to use it or not, we chose a full-factorial experiment design taking into account the effects that written exercises (\textit{WE}), participating in face-to-face exercise meetings (\textit{EM}) and using \naila (\textit{NU}) may have on student performance (\textit{SP}).

We capture different participation levels and  exposure to exercises via a student survey at the end of the semester -- but crucially before the final exam to avoid tweaked answers due to poor exam performance -- by asking the following two questions:

\begin{tcolorbox}[colback=yellow!20, colframe=black, title= \textit{WE}: Written Exercises (4-point scale)]
% \begin{itemize}
% [label=\footnotesize{\textbf{\arabic*}.}]
\textit{How many exercises sheets did you work on?}
\begin{itemize}
    \item 0: None (0 sheets)
    \item 1: Few (1-4 sheets)
    \item 2: Many (5-8 sheets)
    \item 3: (Almost) all (9-10 sheets)
    \end{itemize}
\end{tcolorbox}

\begin{tcolorbox}[colback=yellow!20, colframe=black, title= \textit{EM}: Exercise Meetings (4-point scale)]
% \begin{itemize}
% [label=\footnotesize{\textbf{\arabic*}.}]
\textit{How often did you attend the face-2-face exercise meetings?}
\begin{itemize}
    \item 0: Never (did not participate in any meeting)
    \item 1: Seldom (participated in up to 3 meetings)
    \item 2: Often (missed 2 or 3 meetings)
    \item 3: (Almost) always (missed at most one meeting)
    \end{itemize}
\end{tcolorbox}

More concretely, we base our experimental design on the \textit{Dose-Response Model}, which in our case allows us to determine how students respond to different levels of exposure to our teaching methods (\textit{WE, EM, NU}).
% This model allows us to see whether "more is better" or if there is a "point of diminishing returns", where using "too much" stops helping.
% To honestly evaluate the effect of human or AI feedback to exercises, we only analyzed the population of students that actually performed written exercises ($WE > 0$), as the rest of students ($WE = 0$) is fundamentally different as they have disengaged entirely.
% If you absolutely must keep the WE = 0 students in your dataset to report on the whole class, you must stop treating WE as a continuous linear variable ($0, 1, 2, 3$).The Logic: The jump from "None" (0) to "Few" (1) is a massive behavioral hurdle (going from 0% effort to some effort). The jump from "Few" (1) to "Many" (2) is just a matter of degree.+1The Fix: Convert WE into a factor (categorical variable) in R.Why it works: This allows the regression to calculate a separate baseline intercept for the "None" group, effectively quarantining their 0-scores from the rest of the NAILA analysis. The model will calculate the effect of NU only within the boundaries of the students who cleared that initial hurdle.

To address prior knowledge as a confounding effect, we differentiate \naila's actual impact from the student's baseline ability.
We introduce Baseline Ability (\textit{BA}) as a control variable, which is determined by a student's performance on a 20-minute multiple-choice test about SE basics at the beginning of the semester.

We use the variables \textit{WE}, \textit{EM} and \textit{NU} together with the exam score \textit{SP} and the control variable \textit{BA} to perform a multiple linear regression analysis as shown in the following regression function\footnote{Missing values are treated as NaN.}:

\begin{tcolorbox}[colback=gray!20, colframe=black, title= Multilinear regression for RQ4]
$$SP = \beta_0 + \beta_1 \cdot BA + \beta_2 \cdot WE  + \beta_3 \cdot EM +\:$$
$$\beta_4 \cdot NUC + \beta_5 \cdot NUR$$

\end{tcolorbox}

\begin{figure}[htb]
  \centering
   \includegraphics[width=.5\linewidth]{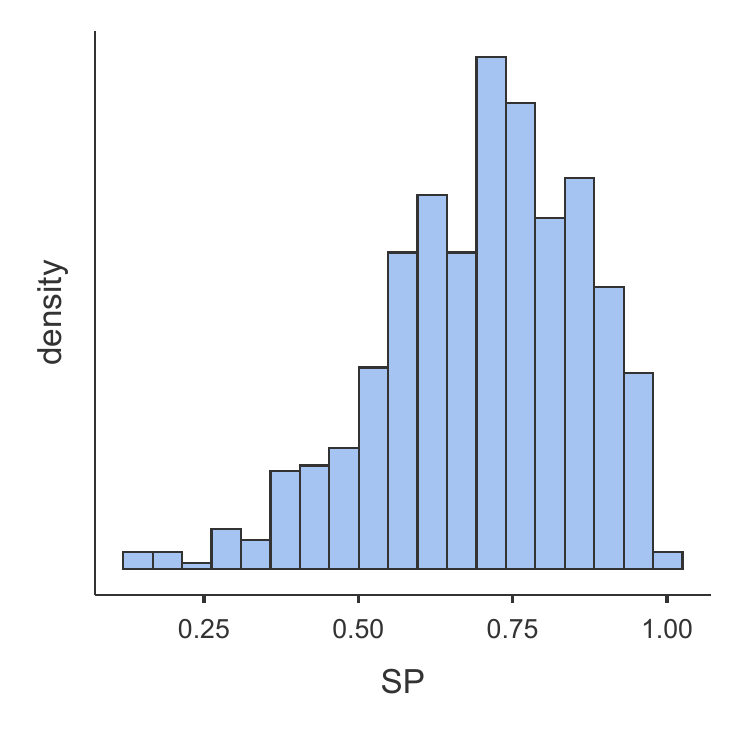}

    \caption{Histogram of Students' Performance (\textit{SP})\\ ($N = 670, \bar{SP} = 70\%, SP_\text{min} = 13\%, SP_\text{max} = 99\%$)}
    \label{fig:SP-hist}
\end{figure}

\textbf{Results RQ4:}
A total of 670 students participated in the final exam.
Figure~\ref{fig:SP-hist} shows the distribution of their performance (\textit{SP}).
The distribution centers around a relatively high median of ca. 72\% indicating students performed rather well. 

\begin{figure}[htb]
  \centering
   \includegraphics[width=1\linewidth]{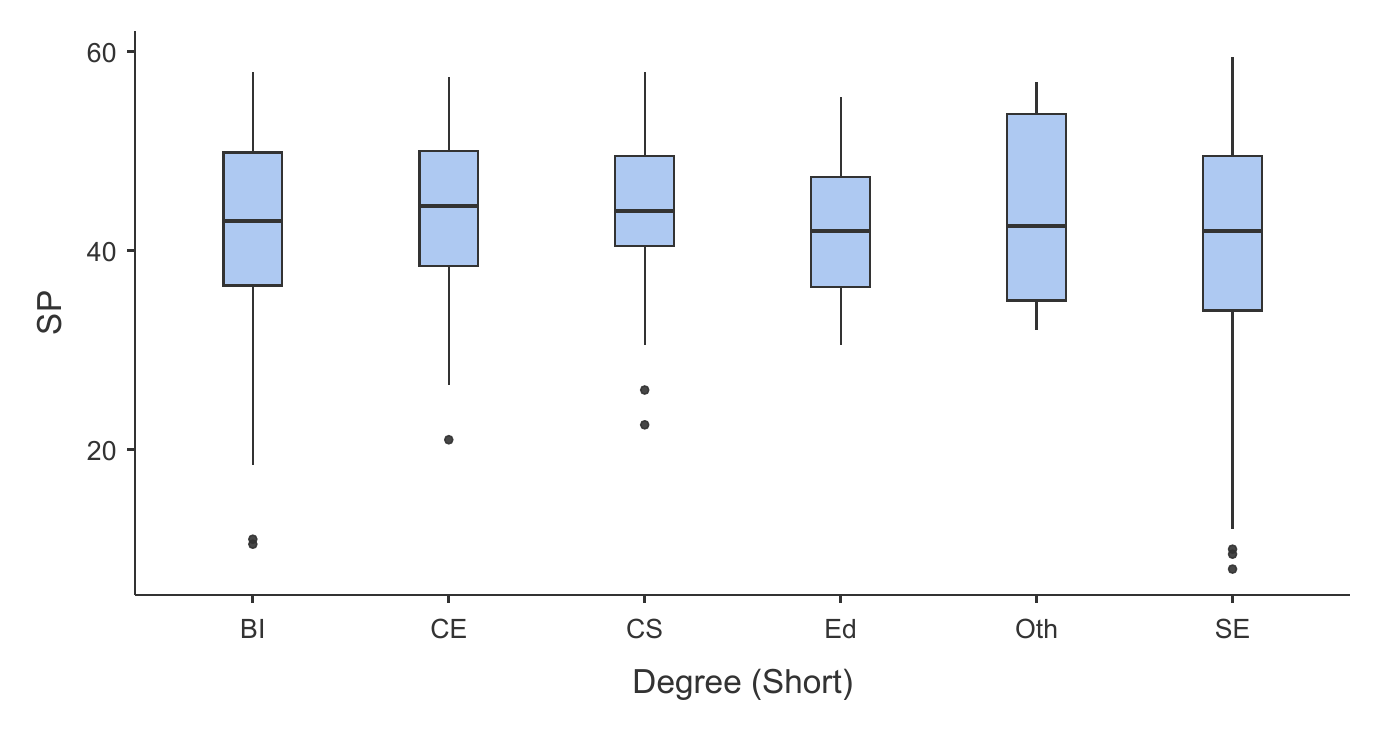}

    \caption{Boxplots of Students' Performance (\textit{SP})\\ ($N = 760$)}
    \label{fig:SP-degree}
\end{figure}

Figure~\ref{fig:SP-degree} shows the box plots for \textit{SP} grouped by degree program, which show visible overlaps and similar median values.
We thus used the non-parametric Kruskal-Wallis test to  confirm that there is no statistical evidence to say that a student's degree has a meaningful impact on the SP score.
We used Kruskal-Wallis, because our data violates the normality assumption (i.e., Shapiro-Wilk test results were all $p < 0.05$ except for CS and Ed).
The result of the Kruskal-Wallis test was $p = 0.08299 > 0.05$ and thus we fail to reject the null hypothesis that there is a difference.
% "Shapiro-Wilk Test Results (Normality):"
%   D       p_value
% 1 BI    0.0000812
% 2 CE    0.00785  
% 3 CS    0.119    
% 4 Ed    0.493    
% 5 Oth   0.0434   
% 6 SE    0.0000384

Table~\ref{tab:WEEM} shows the results for the WE and ME metrics.
As can be seen, while only ca. 6\% of the students did not attempt any written exercise (\textit{WE}), almost 30\% did not participate in any exercise meeting (\textit{EM}).

\begin{table}[bth]
\centering
\begin{tabular}{c|cc|cc}
\hline
\textbf{Level} & \textbf{WE} & \textbf{} & \textbf{EM} & \textbf{} \\ \hline
0 & 39 & 5,8\% & 199 & 29,7\% \\ 
1 & 114 & 17,0\% & 200 & 29,9\% \\
2 & 176 & 26,3\% & 158 & 23,6\% \\
3 & 341 & 50,9\% & 113 & 16,9\% \\ \hline
\end{tabular}
\caption{Distribution by Level ($N = 670$)}
\label{tab:WEEM}
\end{table}

To compute the estimates for the regression coefficients ($\beta_x$), we used the statistics package R. 
Results are shown in Table~\ref{tab:RQ4-results}.
Overall, these results indicate that student success was overwhelmingly and significantly dictated by their  Baseline Ability (\textit{BA}) with a $\beta_1 = 6.98$***, and sheer effort, represented by the volume of Written Exercises completed (\textit{WE}) with a $\beta_2 = 6.58$***.

\begin{table*}[b]
\centering
\begin{tabular}{llrcrrrc}
\hline
\textbf{Coefficient} & \textbf{Term} & \textbf{Estimate} & \textbf{95\% CI} & \textbf{Std. Error} & \textbf{$t$ value} & \textbf{$p$ value} & \textbf{Significance} \\
\hline
$\beta_0$ & (Intercept) & $18.6437$ & $[15.7400, 21.5473]$ & $1.4790$ & $12.605$ & $< 2 \times 10^{-16}$ & *** \\
$\beta_1$ & BA          & $6.9798$  & $[3.5373, 10.4223]$  & $1.7535$ & $3.981$  & $7.56 \times 10^{-5}$ & *** \\
$\beta_2$ & WE          & $6.5750$  & $[5.7339, 7.4160]$   & $0.4284$ & $15.347$ & $< 2 \times 10^{-16}$ & *** \\
$\beta_3$ & EM          & $0.7646$  & $[-0.1151, 1.6443]$  & $0.4481$ & $1.706$  & $0.0884$              & $^.$   \\
$\beta_4$ & NUC         & $3.5914$  & $[0.8176, 6.3651]$   & $1.4128$ & $2.542$  & $0.0112$              & * \\
$\beta_5$ & NUR         & $-0.9808$ & $[-5.5262, 3.5645]$  & $2.3153$ & $-0.424$ & $0.6720$              &       \\
\hline
\end{tabular}
\caption{Multiple Linear Regression Results for Final Exam Performance \textit{SP}}
\label{tab:RQ4-results}
\end{table*}

Crucially, regarding the effectiveness of \naila, the feedback metric (\textit{NUC}) emerged as a significant positive predictor of final performance with a $\beta_4 = 3.59$*. 
This finding supports the hypothesis that students who use the AI's feedback to confirm a high level of mastery before moving on, perform significantly better, thereby demonstrating the value of the AI as a positive reinforcement mechanism. 
Conversely, the feedback metric (\textit{NUR}), which measures relative learning gains through iterative submissions, was not statistically significant.
Consequently, while there is statistical evidence that \naila feedback successfully functions as a "green light", there is currently no evidence that relying on its remedial feedback for course correction translates into measurable performance improvements.

Interestingly, the participation to exercise meetings (captured by \textit{EM}) exhibited only a small marginal positive trend with very low significance with a $\beta_3 = 0.76^.$.

\subsection{Validity Risks}
\label{sec:validity}
Of the 843 active students, only 165 (20\%) participated in the voluntary survey about the reasons for the usage or  non-usage of \naila.
We therefore cannot conclude whether the findings for RQ1 and RQ2 are representative for the broader student base.

A notable validity risk in answering RQ2 is the inherent limitation of the Technology Acceptance Model (TAM) as a proxy for educational value. 
TAM primarily captures subjective satisfaction and perceived usefulness. 
Consequently, there is a risk of the 'Halo effect,' where students report high Perceived Learning (\textit{PL}) simply because the tool was frictionless or enjoyable to use, conflating software satisfaction with actual cognitive development. 
To mitigate this, our study design intentionally separates subjective confidence (evaluated in RQ2) from objective knowledge retention (evaluated via final exam performance in RQ4). 
We do not claim \textit{PL} equates to actual learning; rather, it measures the students' self-efficacy, which our regression model in RQ4 then tests against objective academic realities.

% One concern in using TAM-based questions (RQ2) is the "Halo" effect.
% This effect means that if students find a learning tool "cool" or very easy to use (\textit{PEOU}), they are statistically more likely to report high usefulness (\textit{PU}). 
% We addressed this by extending the standard TAM model to include a distinct "Perceived Learning" (\textit{PL}) dimension. 
% By conceptually separating \textit{PEOU} from specific self-efficacy questions -- such as feeling more confident explaining SE topics or developing practical skills -- we forced students to evaluate their cognitive shifts independently of the software's user interface.

While we provide clear definitions of what the different exposure levels mean in RQ4, the self-reported answers for \textit{WE} and \textit{EM} may still be subject to recall bias.
Unfortunately, we could not perform an objective verification as we did not have access to exercise download logs on the level of individual students,  or attendance records of the exercise meetings. 

% Due to the short time   between the final exam and the CSEE\&T paper deadline, the exam was corrected automatically using classical code for multiple-choice answers and LLMs for free-text answers (RQ4).
% In addition to grading the individual answers and providing reasons for the grading, we instructed the LLM to provide us with a reliability estimate of how confident the LLM is regarding its grading.
% To ensure the reliability of the overall exam grading, we checked answers that were graded with a reliability level below 80\%.
% Still, there may be the risk that students have learned to "write for the AI" rather than mastering the content. 
% While the inclusion of the \textit{BA} control variable helps mitigate this by anchoring \textit{SP} to prior foundational knowledge, future iterations of our study will utilize double-blind human grading for the final evaluation to completely isolate pedagogical gains from algorithmic grading preferences.

As the use of \naila was optional and therefore students could decide for themselves whether to use it or not, we do not achieve the empirical gold standard of randomized control trials.
% Accordingly, we addressed this concern using the measures described above.
% Yet, while we controlled for prior knowledge via \textit{BA}, we could not fully control for unobserved variables like intrinsic motivation.
To mitigate the lack of randomization and the resulting self-selection bias in RQ4 (i.e., highly motivated students being more likely to adopt \naila), our regression model includes behavioral proxies.
Specifically, the volume of Written Exercises completed (\textit{WE}) and Exercise Meetings attended (\textit{EM}) capture the students' baseline effort and engagement. 
By controlling for these factors alongside baseline ability (\textit{BA}), our model isolates the marginal predictive value of the \naila usage metrics (\textit{NU}), independent of a student's general willingness to put effort into the course.

% Ideally, we would have done propensity score matching on observable covariates (such as degree program or prior GPA if available, BA score) between NAILA users and non-users would substantially improve the causal interpretation of RQ4. 

\section{Related Work}
\label{sec:sota}
GenAI offers important opportunities for educating computer science students.
There are numerous studies that analyze the existing work on GenAI for computer science in general~\cite{SahLII24,RaihanSSZ25,WangL25}
Here, we discuss related work specifically targeted at SE education, considering SE a distinct sub-field of computer science.

A widely used application of AI in SE education is LLM-generated feedback for programming assignments (e.g., see~\cite{Choi025,DickB24,SuleimanAW24,JacobsJ24,GrandelSL25,AhmedSLK25,KoutchemeDSH0A025,Bassner2025}).
While this group of work sheds interesting light on the effectiveness of AI in programming education, the nature of programming assignments differs from SE assignments.
Typically, SE includes many other topics as, for instance, covered by the \href{https://www.computer.org/education/bodies-of-knowledge/software-engineering}{SWEBOK}.
These include foundational principles and techniques, different SE roles (such as requirements engineer, tester, or project manager) and different life-cycle models.
Yet, this use of GenAI has not yet been very much reported.

Liao et al. introduced FeedbackPulse that provides personalized guidance for educators to improve the feedback they give to students~\cite{LiaoJCS24}. 
Thus, while FeedbackPulse is a helpful quality check for educators, unlike \naila, it does not provide autonomous feedback to students themselves.

Sölch et al. assessed how groups of students improved their performance across multiple submission attempts with the feedback provided by the AI tool Athena~\cite{SDK25}. 
While the study indicates an increase in scores across attempts, this may not directly help conclude that this indeed facilitated learning.
In particular, as we discussed above and captured by a differentiation into conformance and remedy metrics \textit{PUC} and \textit{PUR}, the feedback provided by Athena may just have helped students to rephrase and improve their answers without having a learning effect.

Many popular learning (management) platforms offer AI features.
\href{https://acemate.ai/}{Acemate} uses AI to transform static educational materials like lecture slides and textbooks into personalized, adaptive learning experiences. Its core AI features include an AI tutor available in over 50 languages, the automated generation of interactive quizzes and flashcards, and advanced learning analytics to help educators identify student knowledge gaps.
\href{https://www.edtech.tum.de/artemis/}{Artemis} features an AI-driven virtual tutor named "Iris," which acts as a context-aware assistant primarily for computer science students. By analyzing problem statements, student code, and error logs, Iris provides calibrated assistance—offering subtle hints and counter-questions rather than full solutions—to foster independent problem-solving skills.
\href{https://www.coursera.org/}{Coursera} utilizes an AI-powered assistant called "Coursera Coach" to give learners on-demand help, such as concept summaries, tailored practice questions, and targeted career guidance. Additionally, it allows instructors to build immersive, interactive learning modules like AI-driven Socratic dialogues that are strictly grounded in the course's expert content.
\href{https://jack-community.org/}{JACK} is an automated assessment system that leverages machine learning and natural language processing to automatically grade complex student assignments, including code and short textual answers. Its AI components provide students with immediate, formative textual feedback while they work, significantly reducing the manual grading burden on educators.
\href{https://moodle.com/}{Moodle} includes a flexible, native AI subsystem that allows institutions to securely plug in their preferred AI models (such as OpenAI, Azure, or open-source local models). Through this integration, users can generate text and images, summarize course content, and explain complex concepts directly within Moodle's standard text editor.
While several of those AI features are similar to \naila, we are not aware of empirical studies that would analyze the effectiveness of these AI features.
 
\section{Discussion}
\label{sec:discussion}

\subsection{Addressing the Challenges}
Referring to  Section~\ref{sec:intro} we can conclude:

\begin{itemize}
    \item \textbf{(a) High number of students} is addressed by introducing \naila, an autonomous tool that scales to provide 24/7 personalized feedback. 
    Findings from RQ3 indicate this scalable approach successfully reached 37\% of the active student body. 
    Furthermore, RQ4 results show that students using this automated feedback to validate their mastery, achieved significantly better final exam performance, effectively mitigating the bottleneck of providing manual feedback for large student cohorts.

    \item \textbf{(b) Diverse types of students} are accommodated by having \naila provide personalized feedback. 
    Results from RQ1 highlight this adaptability, showing students utilized the tool for diverse motivations such as exam preparation (60.9\%) and deep understanding (31.5\%). 
    Additionally, RQ2 findings revealed a high "Perceived Ease of Use" score (4.1 out of 5), suggesting the tool is highly accessible to learners regardless of their prior academic background.

    \item \textbf{(c) Widespread use of GenAI} is supported by findings from RQ2, which confirm that students perceived this AI-driven feedback as highly useful and beneficial for their learning progress. 
    However, RQ4 also cautions against surface-level AI reliance, finding that students who repeatedly adjusted answers just to satisfy the AI's remedial feedback did not exhibit measurable performance improvements.

    \item \textbf{(d) Preference for E-Learning} is addressed by \naila as an easily accessible e-learning resource, thereby directly capturing the 72\% of students who reported favoring AI-based interactive learning. 
    Reflecting this preference, RQ3 found robust digital engagement, with nearly half of \naila users actively utilizing both confirmatory and remedial digital feedback loops. 
    Furthermore, RQ4 results underscore the viability of this e-learning approach, noting that attendance at traditional face-to-face exercise meetings showed only a marginal, low-significance impact on final exam performance.
\end{itemize}

\subsection{Future Work}
Our findings provide a nuanced understanding of how students interact with AI-driven feedback tools, as exemplified by \naila. 
The significant, positive impact of the confirmation metric (\textit{NUC}) suggests that \naila is highly effective as a "green light" mechanism. 
When students utilize the AI to set a high benchmark for themselves—persisting until they achieve a high initial score before moving on—they are actively engaging in self-regulated learning. 
This validation builds confidence and clearly translates to higher overall course performance.

However, the lack of statistical significance for the remedial metric (\textit{NUR}) presents a critical challenge regarding how students utilize remedial feedback. 
While students may improve their scores on a specific exercise through iterative submissions, this localized improvement did not systematically translate to improved course performance. 
One likely explanation is surface-level compliance: students may be using a trial-and-error approach, adjusting their answers to satisfy the AI’s immediate requirements without deeply internalizing the underlying concepts. 
In this scenario, the AI acts more as an answer-checker than a tutor, allowing students to artificially inflate their exercise scores without building the foundational knowledge required for the final exam.

As future work, we thus plan to focus on shifting \naila from a "summative grader" to a "formative coach". 
To prevent students from "gaming" the system through rapid resubmissions, the AI's feedback mechanism could be adjusted to provide strategic hints or Socratic questioning rather than direct corrections. 

\subsection{Regulatory Considerations}
Implementing and deploying GenAI in an educational context requires a structured approach to carefully address relevant regulatory aspects.
In our case, these include the following.

\textit{Copyright:} Students are the authors of their own texts, provided these reach a certain threshold of originality, meaning they go beyond the purely mechanical reproduction of knowledge.
For a text to be protected by copyright, it must be a "personal intellectual creation" (cf. Sec.~2, §2 of the German Copyright Act -- UrhG). 
This requires a certain scope and degree of freedom for individual expression. 
In the written exercises -- at least in our introductory course--, this is was  lacking due to the prescribed structure of answers (e.g., "fill in the blanks"), the mere recall of facts, definitions, or terms from the lecture, and the brevity and conciseness of answers.

Even though the students' answers simply do not reach the required threshold of originality, we were careful to inform the students that their answers will be uploaded and reproduced in the Google Cloud and the Google Gemini AI and that when they upload their exercises to \naila they consent to (1) the reproduction of their work as part of the upload process, (2) the storage and processing of their work, and (3) the use of their work for analysis and evaluation.

\textit{Right of personality:}
Even if copyright is not  applicable, another protected legal right remains that can be even stricter: the general right of personality.
Even without "artistic value," everyone has the right to determine how their personal expressions (which include exercise answers) are used.
As soon as answers can be attributed to a person, it becomes part of their privacy.
Also, publishing a (potentially incorrect) answer without consent can violate the right of personality, as it may portray the student in a potentially negative light.

In our case, we ensured that answers cannot be traced to students (except by the teacher and student itself) by (1) using technical safeguards, particularly pseudonymization (where name and student ID numbers are removed before being sent to the AI), and (2) credentials such that feedback can only be downloaded by the teacher or the student itself.

\textit{Data Protection:} We addressed the applicable data protection requirements under the GDPR as follows: 

(1) As the exercises are voluntary and generating the feedback involves data processing by third-party AI systems, the primary legal basis is explicit consent (Art. 6(1)(a) GDPR).
To this end, we informed students transparently in advance (see above under Copyright).

(2) The student answers -- even if submitted in a voluntary exercise -- are considered  personal data. 
They reflect the student's current level of knowledge, thought processes, and individual writing style.
Students thus maintain their full data protection rights regarding these exercises, including: 
(a) Students have the right to request a copy of their submitted work and the corresponding feedback (Art. 15 GDPR) -- which of course they get from \naila; 
(b) Institutions must clearly communicate how long the submissions will be stored and who has access to them -- we do not store any submissions, after they have been downloaded and once the Web session is closed they are deleted;
(c) Instructors are required to protect student data from unauthorized access -- which we do by running the \naila app and the AI on trusted and protected Google paid services; 
(d) Institutions must use a contractually secured AI solution that explicitly guarantees the input data will not be used to train the provider's AI models and will be deleted after processing -- which is contractually guaranteed as we use Google paid services\footnote{\url{https://ai.google.dev/gemini-api/terms}}.

\textit{AI Act:} 
According to the (current version\footnote{The European Commission introduced the Digital Omnibus package in November 2025, which is a major legislative initiative designed to streamline the EU's increasingly complex digital rulebook, including the GDPR, the Data Act, and cybersecurity frameworks.} of the) EU AI Act, AI systems used in the educational sector to evaluate learning outcomes -- even for formative or voluntary exercises -- are generally classified as High-Risk AI Systems. 
Below we discuss how we addressed the additional applicable requirements of the AI Act (on top of the GDPR):

(1) Students must be informed \textit{in advance} that an AI will be used to generate feedback on their submissions.
In data protection law (Art. 13 GDPR) and under the AI Act, this means at the time of data collection, i.e., at the exact moment the data is collected. 
This is exactly what we do as described above under Copyright.

(2) Even for voluntary feedback, the system must not hallucinate or systematically disadvantage specific writing styles (e.g., those of non-native speakers). 
To address hallucinations, we performed a state-of-the-art parametrization of the LLM by setting temp = 0 and top\_p = 0.
We explicitly prompted the LLM to ignore writing styles, grammatical mistakes and typos.
Finally, we performed a systematic testing of \naila and also incorporated student responses to improve the feedback quality of the system.

(3) If AI feedback is provided, the reasoning behind it should be understandable. 
The AI Act obligates developers of High-Risk AI to build systems where the decision-making pathways are comprehensible to the user (Explainability). 
This is essential for students to actually learn from the feedback and for instructors to exercise appropriate oversight.
We prompted \naila to provide clear explanations of the feedback and the reasons for not giving the maximum points.

\section{Conclusion}
\label{sec:conc}
We introduced \naila, an application that utilizes Large Language Models to provide students with 24/7 autonomous, personalized feedback on their written exercises. 
We conducted an empirical study involving over 900 active students to evaluate why students use \naila, how they accept it, and its ultimate impact on their academic performance. 
The results show that students highly value the tool for exam preparation, and those who used \naila's feedback to confirm their mastery of a topic, achieved significantly higher final exam scores. 
However, students who relied on the AI purely for remedial correction through trial-and-error did not see measurable performance improvements, leading us to conclude that future versions AI feedback must act more like a formative coach than a summative grader, i.e., \naila should rather provide strategic hints and/or Socratic guidance. 

\section*{Acknowledgments and AI Transparency}
I like to thank Sheila Clement, our student coordinator, for providing me with the registration numbers and student statistics, as well as all my my students who participated in the surveys and provided very helpful and constructive feedback on their use of \naila.  The design, execution and evaluation of the empirical study was assisted by the Gemini 3.0 AI. 

\balance

\bibliographystyle{abbrv}
\bibliography{refs,taas}

% \newpage
% \appendix
% \section*{A.1 TAM Breakdown for RQ2}
% \label{sec:A1}

\end{document}